\overfullrule=0pt
\input harvmac
\input amssym.tex
\def\a{{\alpha}}

\def\l{{\lambda}}

\def\b{{\beta}}

\def\g{{\gamma}}

\def\d{{\delta}}

\def\s{{\sigma}}

\def\r{{\rho}}

\def\N{{\nabla}}
\def\Nb{{\overline\nabla}}

\def\O{{\Omega}}
\def\OBL{{\bf\Omega}}
\def\Ob{{\overline\O}}

\def\o{{\omega}}

\def\p{{\partial}}
\def\pb{{\overline\partial}}
\def\t{{\theta}}

\def\L{{\Lambda}}

\def\Pib{{\overline\Pi}}
\def\Jb{{\overline J}}

\def\Tb{{\overline T}}

\def\AB{{\bf A}}
\def\WB{{\bf W}}
\def\FB{{\bf F}}
\def\Wb{{\overline W}}

\baselineskip 12pt

\Title{ \vbox{\baselineskip 12pt
}}
{\vbox{\centerline
{ General fluctuations of the heterotic pure spinor   }
\bigskip
\centerline{ string on curved backgrounds }
}}
\smallskip
\centerline{Osvaldo Chandia\foot{e-mail: ochandiaq@gmail.com}, }
\smallskip
\centerline{\it Departamento de Ciencias, Facultad de Artes Liberales, Universidad Adolfo Ib\'a\~nez}
\centerline{\it \& UAI Physics Center, Universidad Adolfo Ib\'a\~nez}
\centerline{\it Diagonal Las Torres 2640, Pe\~nalol\'en, Santiago, Chile} 

\bigskip
\noindent
The general fluctuations, in the form of vertex operators, for the heterotic superstring in the pure spinor formalism are considered. The case of a flat background is reviewed and the case of a curved case is studied. The left-moving ghosts are pure spinors and the right-moving ghosts come from fixing the reparametrization invariance. The role of the reparametrization ghosts in the construction of the vertex operators is emphasized. 

\Date{February 2019}


\lref\BerkovitsFE{
  N.~Berkovits,
  ``Super Poincare covariant quantization of the superstring,''
JHEP {\bf 0004}, 018 (2000).
[hep-th/0001035].
}

\lref\BerkovitsNN{
  N.~Berkovits,
  ``Cohomology in the pure spinor formalism for the superstring,''
JHEP {\bf 0009}, 046 (2000).
[hep-th/0006003].
}

\lref\BerkovitsMX{
  N.~Berkovits and O.~Chandia,
  ``Lorentz invariance of the pure spinor BRST cohomology for the superstring,''
Phys.\ Lett.\ B {\bf 514}, 394 (2001).
[hep-th/0105149].
}

\lref\BerkovitsQX{
  N.~Berkovits and O.~Chandia,
  ``Massive superstring vertex operator in D = 10 superspace,''
JHEP {\bf 0208}, 040 (2002).
[hep-th/0204121].
}

\lref\BerkovitsPH{
  N.~Berkovits and B.~C.~Vallilo,
  ``Consistency of superPoincare covariant superstring tree amplitudes,''
JHEP {\bf 0007}, 015 (2000).
[hep-th/0004171].
}

\lref\BerkovitsPX{
  N.~Berkovits,
 ``Multiloop amplitudes and vanishing theorems using the pure spinor formalism for the superstring,''
JHEP {\bf 0409}, 047 (2004).
[hep-th/0406055].
}

\lref\WittenNT{
  E.~Witten,
  ``Twistor - Like Transform in Ten-Dimensions,''
Nucl.\ Phys.\ B {\bf 266}, 245 (1986).
}

\lref\BerkovitsBT{
  N.~Berkovits,
 ``Pure spinor formalism as an N=2 topological string,''
JHEP {\bf 0510}, 089 (2005).
[hep-th/0509120].
}

\lref\LeeJY{
  K.~Lee and W.~Siegel,
  ``Conquest of the ghost pyramid of the superstring,''
JHEP {\bf 0508}, 102 (2005).
[hep-th/0506198].
}

\lref\BerkovitsUE{
  N.~Berkovits and P.~S.~Howe,
  ``Ten-dimensional supergravity constraints from the pure spinor formalism for the superstring,''
Nucl.\ Phys.\ B {\bf 635}, 75 (2002).
[hep-th/0112160].
}

\lref\ChandiaIX{
  O.~Chandia,
  ``A Note on the classical BRST symmetry of the pure spinor string in a curved background,''
JHEP {\bf 0607}, 019 (2006).
[hep-th/0604115].
}

\lref\ChandiaAFC{
  O.~Chandia and B.~C.~Vallilo,
  ``A superfield realization of the integrated vertex operator in an $AdS_5\times S^5$ background,''
JHEP {\bf 1710}, 178 (2017).
[arXiv:1709.05517 [hep-th]].
}

\lref\ChandiaEKC{
  O.~Chandia and B.~C.~Vallilo,
  ``Vertex operators for the plane wave pure spinor string,''
JHEP {\bf 1810}, 088 (2018).
[arXiv:1807.05149 [hep-th]].
}

\lref\ChandiaHN{
  O.~Chandia and B.~C.~Vallilo,
  ``Conformal invariance of the pure spinor superstring in a curved background,''
JHEP {\bf 0404}, 041 (2004).
[hep-th/0401226].
}

\lref\ChandiaKJA{
  O.~Chandia, A.~Mikhailov and B.~C.~Vallilo,
 ``A construction of integrated vertex operator in the pure spinor sigma-model in $AdS_5 \times S^5$,''
JHEP {\bf 1311}, 124 (2013).
[arXiv:1306.0145 [hep-th]].
}

\lref\BerkovitsYR{
  N.~Berkovits and O.~Chandia,
  ``Superstring vertex operators in an $AdS_5\times S^5$ background,''
Nucl.\ Phys.\ B {\bf 596}, 185 (2001).
[hep-th/0009168].
}

\newsec{Introduction}

One possible description of the fluctuations around classical configurations in string theory is in terms of vertex operators. They have two properties. The first is that the set of vertex operators describe the physical content of the string theory in consideration. The second is that they can be used as asymptotic states in scattering amplitudes. In all string theories, there is a common form to define vertex operators. They are in the cohomology of a BRST charge. For the bosonic string, the BRST charge is given by gauge fixing the conformal symmetry of the world-sheet theory. For the RNS string, the BRST charge is given by gauge fixing the superconformal symmetry of the world-sheet theory. If the background space-time contains Ramond states, the RNS string weakens its power to perform computations such that scattering amplitudes and loses manifest background isometries. For example, space-time supersymmetry is not manifest in flat ten-dimensional space-time. A world-sheet string model that avoids this problem is the pure spinor string  \BerkovitsFE.  Although it is not known the symmetry to gauge-fix and get a BRST operator, it is known the existence of a nilpotent charge that gives the correct cohomology \BerkovitsNN\BerkovitsMX\BerkovitsQX\ and define a correct scattering amplitudes in flat space-time \BerkovitsPH\BerkovitsPX. The form of this charge is simple and it is equal to $Q=\oint \l^\a d_\a$, where $d_\a$ is the world-sheet generator of superspace translations and $\l$ is a pure spinor variable, that is, it satisfies $\l\g^m\l=0$ with $\g^m$ being the symmetric $16\times 16$ gamma matrices in ten dimensions. Consider the open string in a flat background. In the massless sector, the vertex operator is the ghost number one  operator $U=\l^\a A_\a$, where $A_\a$ is a function of $N=1$ ten-dimensional superspace. Imposing $U$ to be in the cohomology of $Q$ implies that the superfield $A_\a$ satisfies the equation $\g_{mnpqr}^{\a\b}D_\a A_\b=0$ and is defined up to the gauge invariance $\d A_\a=D_\a\O$, where $D_\a$ is the superspace covariant derivative and $\O$ is a scalar superfield. The equation and the gauge invariance of $A_\a$ state that this superfield has the physical degrees of freedom corresponding to the photon and the photino \WittenNT. To compute scattering amplitudes it is necessary the knowledge of the integrated vertex operators. It is defined through the relation $QV=\p U$. In this case, the integrated vertex operator is a conformal weight one field and it is expressed in terms of higher fermionic derivatives of $A_\a$, as we review below. In this construction there will appear superfields like the vector potential $A_m$ (whose $\t$-independent component is the photon) and the fermionic field-strength $W^\a$ (whose $\t$-independent component is the photino).  The integrated vertex operator can be used to deform the world-sheet action. The deformed action preserves the symmetries of the classical action (namely, BRST and conformal symmetries) and a new BRST invariant system is obtained. A natural question is how to generalize this discussion, in particular the deformed action, in a non-flat background.  The first step to study the vertex operators of a world-sheet system in a curved background. This has been done for the $AdS_5\times S^5$ background in the pure spinor formalism. The unintegrated vertex was obtained in \BerkovitsYR\ and the integrated vertex operator was studied in \ChandiaKJA\ and \ChandiaAFC. What about other backgrounds or even a generic supergravity background? We answer this question for the case of the heterotic string in a generic supergravity background.

It is interesting to note that in the bosonic string the relation between integrated and unintegrated is straightforward because the existence of the parametrization ghosts $b$ and $c$. The integrated vertex operator is $V=\p X^m A_m(X)$ and the unintegrated vertex is $U=c\p X^m A_m(x)$ and the relation $QV=\p U$ is satisfied in the Lorentz gauge. In this case, we simply have $U=cV$. In the pure spinor version of the open string, the reparametrizaton ghost are not fundamental fields and they can be constructed from the fundamental fields of the pure spinor string \BerkovitsBT. However, there exist fundamental $b, c$ ghosts in the heterotic string as the anti-holomorphic sector is described in terms of the bosonic string. We study the implications that the existence of these ghosts have in the construction of vertex operators in the case of the heterotic string in a generic background.

We review the flat ten-dimensional space-time case in section 2. The unintegrated vertex for generic supergravity and super Yang-Mills fluctuations are studied in section 3. Here, vector potential and fermionic field-strength superfields are defined. The next step is to construct the integrated vertex in curved background, we perform such construction in section 4. We end with concluding remarks in section 5.

\subsec{Conventions}

The background superspace coordinates in ten dimensions are denoted by $Z^M=(X^m,\t^\mu)$ with $m=0, \dots, 9$ and $\mu=1, \dots 16$. There exists local flat coordinates denoted by $Z^A=(X^a,\t^\a)$ with $a=0, \dots 9$ and $\a=1, \dots, 16$. The relation between these coordinates is $Z^M=Z^A E_A{}^M(Z)$, where $E_A{}^M$ is the vielbein superfield and its inverse is $E_M{}^A$. We use $M, N, \dots=(m, \mu), (n, \nu), \dots$ are the curved superspace indices and $A, B, \dots=(a, \a), (b, \b), \dots$ are local superspace indices. Note that in a flat background there is no distinction between curved and local indices. 

\newsec{Review of the heterotic string in a flat background}

We now review the heterotic string in a flat ten-dimensional background using the pure spinor formalism. The action is given by
\eqn\shetz{ S= \int d^2z ~ \ha\p X^m \pb X_m + p_\a \pb \t^\a +\o_\a \pb \l^\a + b \p c + \ha \r_{\cal A} \p \r_{\cal A} ,}
where $(X,\t)$ are the superspace coordinates in flat ten-dimensional space-time (that is, $m, n, \dots$ run from $0$ to $9$ and $\a, \b, \dots$ run from $1$ to $16$), $p$ is the momentum conjugate of $\t$, $(\l,\o)$ are the pure spinor conjugate variables, $(b, c)$ are the right-moving reparametrization ghosts and $\r$ is the heterotic fermion which live in a representation of the group $SO(32)$ or the $SO(16)\times SO(16)$ subgroup of $E_8\times E_8$  such that ${\cal A}, {\cal B}, \dots$ run from $1$ to $32$. The pure spinor variables are constrained by the condition $\l\g^m\l=0$ and the invariance $\d\o_\a=(\l\g^m)_\a\L_m$, then only $22$ out of $32$ $\l$ and $\o$ variables are independent \BerkovitsFE. 

Physical states are given in the cohomology of the BRST operator
\eqn\Qz{Q=\oint \l^\a d_\a + c\Tb + bc\pb c ,}
where $d_\a$ is the generator of superspace translations and $\Tb$ is the right moving stress-energy tensor for the $(X,\r)$ system. Note that $Q^2=0$ provided the pure spinor condition and that the OPE $\Tb(\bar y)\Tb(\bar z)$ has central charge $26$ ($10$ from $X$ and $16$ from the heterotic fermions). The massless unintegrated vertex operator has $\l$-ghost number $1$ and $(bc)$-ghost number $1$ and it is given by
\eqn\Uz{ U = \l^\a \AB_{\a m} ~c\pb X^m + \l^\a \AB_{\a I} ~c\Jb^I ,}
where $\Jb^I = \ha K^I_{\cal AB} \r_{\cal A} \r_{\cal B}$ with $K^I$ being the Lie generators of the $E_8\times E_8$ or $SO(32)$ groups in the heterotic fermions representation used above. This means that these generators satisfy the algebra $[K_I,K_J]=f_{IJ}{}^K K_K$, where $f_{IJ}{}^K$ are the structure constants of  the $E_8\times E_8$ or $SO(32)$ groups ($I=1, \dots, 496$). Note that we use boldface symbols for the fluctuations.  

Both terms in \Uz\ are independently annihilated by the BRST charge. Consider the action of $Q$ on the first term in \Uz. Note that the action of the right-moving variables put the superfield $\AB_{\a m}$ satisfying the gauge fixing condition $\p^m \AB_{\a m}=0$ and the equation $\p^n\p_n \AB_{\a m}=0$. Although there exists a way to overcome this lack of covariance (see \LeeJY) we will focus on the covariant equations implied by action of the the pure spinor part of \Qz. It implies that
\eqn\DUz{ \l^\a\l^\b D_\a \AB_{\b m}=0\Rightarrow D_{(\a} \AB_{\b)m} = \g^n_{\a\b}\AB_{nm} ,}
where $D_\a$ is the ten-dimensional covariant superspace derivative. 

As it is well known \WittenNT, the equation in \DUz\ implies the existence of a superfield $\WB^\b{}_m$ satisfying
\eqn\Wz{ D_\a \AB_{nm} - \p_n \AB_{\a m} = (\g_n)_{\a\b} \WB^\b{}_m .}
The superfield $\WB$ has fermionic derivative equal to
\eqn\DWz{ D_\a \WB^\b{}_m = {1\over4} (\g^{pn})_\a{}^\b \FB_{pnm} ,}
where $\FB_{pnm}=\p_{[p}\AB_{n]m}$.  

In summary, the equation \DUz\ allows to have the chain of relations
\eqn\gravity{\eqalign{&D_{(\a}\AB_{\b)m}=\g^n_{\a\b}\AB_{nm},\cr &D_\a \AB_{nm}-\p_n\AB_{\a m}=(\g_n)_{\a\b}\WB^\b{}_m,\cr &D_\a \WB^\b{}_m={1\over4}(\g^{pn})_\a{}^\b \FB_{pnm} .}}
The equations \gravity\ imply that the $\t$-independent term in $\AB_{nm}$ and $\WB^\a{}_m$ describe the massless fields of supergravity in ten-dimensions (around flat space-time). In fact, these superfields satisfy the equations
\eqn\eqz{ \p^p \FB_{pnm}=\g^n_{\a\b} \p_n \WB^\b{}_m=0.}
Similarly for the second term in \Uz\ we have the equations
\eqn\gauge{\eqalign{&D_{(\a} \AB_{\b)I}=\g^n_{\a\b} \AB_{nI},\cr
&D_\a \AB_{nI} - \p_b \AB_{\a I} = (\g_n)_{\a\b} \WB^\b{}_I , \cr
&D_\a \WB^\b{}_I={1\over4} \FB_{pnI} , }}
where $\FB_{pnI}=\p_{[pI}\AB_{n]I}$. The equations \gauge\ imply that the $\t$-independent term in $\AB_{nI}$ and $\WB^\a{}_I$ describe the massless fields of SYM in ten-dimensions (around flat space-time). In fact, these superfields satisfy the equations
\eqn\eqzI{ \p^p \FB_{pnI}=\g^n_{\a\b} \p_n \WB^\b{}_I=0.}

The gauge invariance of the background superfields are obtained as BRST invariance of the unintegrated vertex operator. That is, $U$ of \Uz\ is defined up to $Q(\OBL_m c\pb X^m +\OBL_I c\Jb^I )$ for some superfields $\OBL_m$ and $\OBL_I$. Then, the gauge transformations for the $\AB$ in \Uz\ are
\eqn\gUz{\d \AB_{\a m}=D_\a \OBL_m,\quad \d \AB_{Im}=D_\a\OBL_I .}
They imply that the $\AB_{nm}$ and $\AB_{nI}$ transform as vector potentials, that is
\eqn\gpotz{\d \AB_{nm}=\p_n\OBL_{m},\quad \AB_{nI}=\p_n\OBL_I ,}
and the superfields $\WB$ and $\FB$ are gauge invariant. This shows that the physical content in \Uz\ are the supergravity and SYM fluctuations around flat space in ten dimensions. 

We now review the construction of the integrated version of \Uz\ \BerkovitsUE. The integrated vertex operator $V$ is defined through $QV=\p {\overline W}-\pb W$, where $Q{\overline W}=\pb U$ and $QW=\p U$.  It turns out that
\eqn\WWz{ \eqalign{ &W= (\p\t^\a \AB_{\a m}+\Pi^n \AB_{nm} + d_\a \WB^\a{}_m +\ha N^{pn} \FB_{pnm}) c\pb X^m \cr&~~~~~~ + (\p\t^\a \AB_{\a I}+\Pi^n \AB_{nI} + d_\a \WB^\a{}_I +\ha N^{pn} \FB_{pnI} ) c \Jb^I ,\cr
&{\overline W} = -\l^\a ( \AB_{\a m}\pb X^m + \AB_{\a I} \Jb^I ) ,\cr  }}
where $\Pi^n=\p X^n +\ha(\t\g^n\p\t)$ is the supersymmetric momentum and $N^{pn}=\ha(\l\g^{np}\o)$ is the Lorentz generator for the pure spinor variables. Finally, the integrated vertex operator satisfying $QV=\p{\overline W}-\pb W$ becomes 
\eqn\Vz{ \eqalign{ &V= (\p\t^\a \AB_{\a m}+\Pi^n \AB_{nm} + d_\a \WB^\a{}_m +\ha N^{pn} \FB_{pnm}) \pb X^m \cr&~~~~~~ + (\p\t^\a \AB_{\a I}+\Pi^n \AB_{nI} + d_\a \WB^\a{}_I +\ha N^{pn} \FB_{pnI} )  \Jb^I .\cr }}

The goal of this paper is to generalize these results to an on-shell curved background. Before going into the details of the curved background case, let me point an interesting structure from $U, W, \Wb, V$ implied by the presence of the pair of reparametrization ghosts $(b, c)$. Note that $\Wb$ is related to $U$ and $W$ is related to $V$ according to
\eqn\UcW{U=c\Wb,\quad W=cV .}
Because $V=\{b,W\}$, it can be verified that 
\eqn\QV{QV=[\Tb,W]-[b,QW]=\pb W-[b,c\p \Wb] =\pb W-\p \Wb ,}
as is required. Here we used that $Qb=\Tb$ with $\Tb$ being the right-moving stress-energy tensor for $X$ and the heterotic fermions. This is not a surprise because it is already known that $V$ satisfies \QV, but the argument can be reversed in the following way. The first equation in \UcW\ is a fact and looking at equation \WWz\ one can sees that $W$ has the form $W=cX$. Following \QV\ one can obtain that $X$ satisfies the equation for the integrated vertex. Therefore, $X$ will be the integrated vertex operator and, actually, the expression $\Vz$ is $X$.  We use this analysis below to find the integrated vertex operator in a generic curved background.

\newsec{The unintegrated vertex operator of the heterotic string in a curved background}

The action of the heterotic string in a generic curved background is obtained by adding $\int d^2z V$  to the action \shetz\ and covariantizing respect to the background invariance \BerkovitsUE. The resulting action is
\eqn\Shet{S=\int  d^2z ~\ha\Pi^a\Pib_a+\ha\Pi^A\Pib^BB_{BA} +\o_\a\Nb\l^\a + \ha\r_{\cal A} \N \r_{\cal A} }
$$+d_\a(\Pib^\a+\Jb^I W_I^\a) +\l^\a \o_\b \Jb^I U_{I\a}{}^\b + b\p c + S_{FT},$$
where the background superfields depend on the superspace coordinates $Z^M$. Note that any field $F^A$ with  local superspace indices $A=(a, \a)$ is related to the same field, $F^M$ with curved superspace indices $M=(m, \mu)$ according to $F^A=F^M E_M{}^A$, where $E_M{}^A$ is the vielbein superfield. Also, $\Pi^A=\p Z^M E_M{}^A, \Pib^A=\pb Z^M E_M{}^A$ with $E(Z)$ being the supervielbein. The covariant derivatives in the action are given by
\eqn\covD{\Nb\l^\a = \pb\l^\a+\l^\b\Ob_\b{}^\a,\quad \N\r_{\cal A} =\p\r_{\cal A} + A_I K^I_{\cal AB} \r_{\cal B} ,}
where $\Ob_\b{}^\a=\pb Z^M \O_{M\b}{}^\a$ and $A_I=\p Z^M A_{MI}$ with $\O_M$ being the Lorentz connection and $A_M$ being the gauge potential. The Lorentz connection has the structure \BerkovitsUE\
\eqn\LorConn{\O_{M\a}{}^\b=\O_M\d_\a^\b+{1\over4}(\g^{ab})_\a{}^\b\O_{Mab} ,}
where $\O_{Mab}$ is the usual Lorentz vector connection and $\O_M$ is necessary to preserve conformal invariance of \Shet\ as it was shown in \ChandiaHN.

Note that $\Jb^I=\ha K^I_{\cal{AB}}\r_{\cal A}\r_{\cal B}$ just like in flat space background. The other background superfields in the action are $W_I^\a$ and $U_{I\a}{}^\b$. $W$ contains the gaugino as its lowest $(\t,{\overline\t})$ component and $U$ contains the field-strength gauge boson in its lowest $(\t, {\overline\t})$ component. $S_{FT}$ is the Fradkin-Tseytlin term defined as 
\eqn\sft{S_{FT}=\int d^2z ~ r\Phi ,} 
where $r$ is the world-sheet curvature and $\Phi$ is the dilaton superfield which is related to the first term in \LorConn\ as $4\O_\a=\N_\a\Phi$. 

The action \Shet\ is invariant under the transformations generated by $Q=\int (\l^\a d_\a + c\Tb+bc\p c)$. This charge is nilpotent and conserved when the background satisfies certain constraints. They put the background to satisfy the equations of supergravity and super Yang-Mills in a curved background as it was shown in \BerkovitsUE. These constraints are solved with the torsion and field-strength components
\eqn\torsion{ T_{\a\b}{}^a = -\g^a_{\a\b},\quad T_{A\a}{}^\b = 0,\quad T_{\a a}{}^b=2(\g_a{}^b)_\a{}^\b \O_\b,\quad F_{I\a\b}=0,\quad F_{Ia\a}=-(\g_a)_{\a\b}W^\b_I ,}
where $\O_\a$ is the scalar part of the Lorentz connection and turns out to be proportional to the fermionic derivative of the dilaton superfield. Note that the components of the $3$-form $H=dB$ are constrained by BRST invariance as it was shown in \BerkovitsUE.  For example, $H_{\a\b a}=-(\g_a)_{\a\b}$ and $H_{\a\b\g}=0$. 

The fields in the action transform under the BRST charge $Q$ as \ChandiaIX
\eqn\brst{\eqalign{&Q\Pi^A=\d^A_\a\N\l^\a-\l^\a\Pi^BT_{B\a}{}^A,\quad  Q\l^\a =Q \r_{\cal A} = 0\cr &Qd_\a=-(\l\g_a)_\a \Pi^a +\l^\b\l^\g\o_\d R_{\a\b\g}{}^\d,\quad Q \o_\a =d_\a , }}
where $R=d\O+\O\wedge\O$ is the background curvature superfield. Note that these transformations are given up to a Lorentz and gauge transformation terms with field-dependent parameters (for example, the transformation of $\Pi^\a$ includes a term $\Pi^\b\l^\g\O_{\g\b}{}^\a$ \ChandiaIX) but these contributions are ignored when $Q$ acts on scalars combinations like the Lagrangian above or the vertex operators below. However these contributions are important to verify nilpotence of $Q$ on the world-sheet fields. Note that, the pure spinor BRST charge acts non-trivially on $\Pib^A$. In fact, 
\eqn\Qpib{Q\Pib^A=\d^A_\a\Nb\l^\a-\l^\a\Pib^B T_{B\a}{}^A .}
This is used in verifying that the action is BRST invariant \ChandiaIX\ and it will be used below.

The fluctuations around a generic background are given by vertex operators. The unintegrated vertex is
\eqn\Un{ U = \l^\a \AB_{\a a}~c\Pib^a + \l^\a \AB_{\a I}~c\Jb^I ,}
where the superfield fluctuations depend on the supercoordinates, that is $\AB(Z)$. As in flat space, we use boldface symbols for the fluctuations. Not confuse $\AB$ with the gauge potential $A$ of the action \Shet. The vertex operators are well-defined on-shell objects, that is the equations of motion from the action \Shet\ are imposed. In this case, by varying respect to $d_\a$ we have the equation $\Pib^\a+\Jb^I W_I^\a=0$, therefore a term with $\Pib^\a$ similar to the first term of $U$ is not an independent world-sheet variable and can be moved to the second term in \Un. 

The action of pure spinor BRST charge on the vertex  \Un\ gives the equation
\eqn\QUA{ \l^\a \l^\b \left[ \left( \N_\a \AB_{\b a} + T_{\a a}{}^b \AB_{\b b} \right) c\Pib^a+ \left( \N_\a \AB_{\b I} + \g^a_{\a\g} \AB_{\b a} W^\g_I \right) c\Jb^I \right] =0,}
which implies the existence of the superfields $\AB_{b a}$ and $\AB_{b I}$ as in \DUz\ and \gauge\ through the equations
\eqn\DAs{ \N_{(\a} \AB_{\b) a} + T_{(\a a}{}^b \AB_{\b) b} = \g^b_{\a\b} \AB_{b a},\quad \N_{(\a} \AB_{\b)I} - F_{I(\a}{}^a \AB_{\b) a}  = \g^a_{\a\b} \AB_{a I} .}

As in flat space, we find higher fermionic derivatives. Consider the first equation in \DAs. After using the (anti-)commutation relations and Bianchi identities of the background superspace geometry, the following equation is obtained
\eqn\Dapa{ \g^b_{(\a\b} \left( \N_{\g)} \AB_{ba} - \N_b \AB_{\g)a} - 2 \O_{\g)} \AB_{ba}  + T_{ab}{}^c \AB_{\g)c} + T_{\g)a}{}^c \AB_{bc} \right) =0,}
which implies the existence of a superfield $\WB^\b{}_a$ satisfying  
\eqn\Wc{ \N_{\a} \AB_{ba} - \N_b \AB_{\a a} - 2 \O_{\a} \AB_{ba}  + T_{ab}{}^c \AB_{\a c} + T_{\a a}{}^c \AB_{bc} = (\g_b)_{\a\b} \WB^\b{}_a .} 
This is the analogous to \Wz\ in heterotic curved superspace.

The next step is to find the fermionic covariant derivative of $\WB^\b{}_a$. This is more involving. It turns out that the following combination
\eqn\defP{\eqalign{ {\bf\Psi}_\a{}^\b{}_a &=\N_\a \WB^\b{}_a - 4 \O_\a \WB^\b{}_a +T_{\a a}{}^b \WB^\b{}_b - T_{ab}{}^\b \AB_\a{}^b \cr &+ 2(\g^b)^{\b\s} ( (\N_\a \O_\s) \AB_{ba} - \O_\s \N_b \AB_{\a a} + \O_\s T_{ab}{}^c \AB_{\a c} ) ,}}
satisfies the equation
\eqn\eqP{ \eqalign{10 {\bf\Psi}_\a{}^\b{}_a + \g^b_{\a\r} \g_b^{\b\g} {\bf\Psi}_\g{}^\r{}_a &= (\g^{cb})_\a{}^\b ( \FB_{cba} + T_{cb}{}^A \AB_{A a} - T_{a[c}{}^d \AB_{b]d} +  3\tau_{cb}{}^d \AB_{da} )\cr &- (\g^{edcb})_\a{}^\b \tau_{edc} \AB_{ba} ,} }
where $\FB_{cba}=\N_{[c} \AB_{b]a}$ and $\tau_{abc}=\g_{abc}^{\a\b} \O_\a \O_\b$. To obtain this equation one starts with $\WB^\b{}_a$ from \Wc\ and applies the operator $\N_\a$. It is necessary to use the torsion constraints \torsion\ and Bianchi identities. The equations in section 5 of \ChandiaHN\ are used. The solution of \eqP\ is
\eqn\DWc{ {\bf\Psi}_\a{}^\b{}_a={1\over4}  (\g^{cb})_\a{}^\b ( \FB_{cba} + T_{cb}{}^A \AB_{A a} - T_{a[c}{}^d A_{b]d} + 3\tau_{cb}{}^d \AB_{da} ) - {1\over{12}} (\g^{edcb})_\a{}^\b \tau_{edc} \AB_{ba} .}
 This is the analogous of \Wz\ for heterotic curved background. Note the existence of $4$-form projection.

Similarly, we can find the consequences of the second equation in \DAs. As it was done in \Dapa, we find that
\eqn\Dapi{ \g^a_{(\a\b} \left( \N_{\g)} \AB_{aI} - \N_a \AB_{\g)I} - 2\O_{\g)} \AB_{aI} - \g^b_{\g)\r} W^\r_I \AB_{ab}  + F_{Ia}{}^b \AB_{\g)b} \right) = 0,}
which implies the existence of $\WB^\b{}_I$ satisfying
\eqn\Wci{ \N_{\a} \AB_{aI} - \N_a \AB_{\a I} - 2\O_{\a} \AB_{aI} - \g^b_{\a\r} W^\r_I \AB_{ab}  +F_{Ia}{}^b \AB_{\a b}  = (\g_a)_{\a\b} \WB^\b{}_I .}
This is the analogous of the second equation in \gauge\ in a curved background. The equation for $\N_\a\WB^\b{}_I$ is obtained by noting that
\eqn\defPI{\eqalign{{\bf\Psi}_\a{}^\b{}_I &= \N_\a \WB^\b{}_I-4\O_\a\WB^\b{}_I -\g^a_{\a\s} W^\s_I\WB^\b{}_a + \N^a W^\b_I \AB_{\a a} + f_{IJK} W^\b_J \AB_{\a K} \cr
&+2(\g^a)^{\b\s} \left(\N_\a\O_\s A_{aI} - \O_\s\N_a A_{\a I} +\O_\s F_{Ia}{}^b A_{\a b} \right) ,}}satisfies the equation
\eqn\eqPI{\eqalign{10{\bf\Psi}_\a{}^\b{}_I+\g^a_{\a\r}\g_a^{\b\g}{\bf\Psi}_\g{}^\r{}_I&=(\g^{ab})_\a{}^\b\left(\FB_{Iab}+T_{ab}{}^A \AB_{AI} + \AB_{[a}{}^cF_{Ib]c}+3\tau_{ab}{}^c \AB_{cI} \right)\cr &- (\g^{abcd})_\a{}^\b \tau_{abc}\AB_{dI} ,}}
where $\FB_{Iab}=\N_{[a}\AB_{b]I}$. This equation is solved by 
\eqn\PsI{{\bf\Psi}_\a{}^\b{}_I={1\over4}(\g^{ab})_\a{}^\b\left(\FB_{Iab}+T_{ab}{}^A \AB_{AI} + \AB_{[a}{}^cF_{Ib]c}+3\tau_{ab}{}^c \AB_{cI} \right)-{1\over{12}}(\g^{abcd})_\a{}^\b \tau_{abc}\AB_{dI} ,}
this is the heterotic curved version of third equation in \gauge. Again, there appears a $4$-form projection.

Before ending this section, the equation of motion of the fluctuations should be obtained. That is, the analogous of equations \eqz\ and \eqzI\ in curved heterotic background. It is obtained as follows. Consider \DWc\ and compute $\N_{(\b} \Psi_{\a)}{}^\b{}_a$. First, apply $\N_\b$ (and symmetrize in $(\a,\b)$) into the definition \defP. Second, apply  $\N_\b$ (and symmetrize in $(\a,\b)$) into the solution \DWc\ and then compare the results. The result has the form  
\eqn\NWc{\g^b_{\a\b}\left( \N_b\WB^\b{}_a -(\N_b\Phi)\WB^\b{}_a + T_{ba}{}^c\WB^\b{}_c + (\g^c)^{\b\g}\O_\g \FB_{cba} -2(\g^c)^{\b\g} T_{bc}{}^\r A_{\r a} + \cdots \right) = 0 ,} 
where $\Phi$ is the Fradkin-Tseytlin scalar superfield which determines the scalar connection through $4\O_\a=\N_\a\Phi$. The terms in $\cdots$  depend on the torsion, curvatures and the fluctuations. They are not so illuminating so I have not found them explicitly. The important property is that  this equation has a gamma matrix in the front just like in flat space-time.  A similar equation is obtained for \defPI\ which has the form $\g^a_{\a\b}\N_a\WB^\b{}_I+\cdots=0$. 

Let's discuss the gauge invariance of \Un. As in flat space, the unintegrated vertex operator is defined up to $Q(\OBL_a c\Pib^a+\OBL_I c\Jb^I )$. This implies that,
\eqn\gUc{\d\AB_{\a a}=\N_\a\OBL_a+T_{\a a}{}^b\OBL_b,\quad \d\AB_{\a I}=\N_\a\OBL_I-\g^a_{\a\b}\OBL_a W^\b_I,}
which implies that the superfields in \DAs\ are defined up to
\eqn\gAv{\d\AB_{ba}=\N_b\OBL_a+T_{ba}{}^c\OBL_c,\quad \d\AB_{aI}=\N_a\OBL_I+F_{Ia}{}^b\OBL_b .}
Similarly, the gauge transformation for the gaugino field-strengths are
\eqn\dWs{\eqalign{&\d\WB^\a{}_a=-T_{ab}{}^\a \OBL^b -2(\g^b)^{\a\b}\O_\b(\N_b\OBL_a+T_{ba}{}^c\OBL_c) ,\cr&\d\WB^\a{}_I=f_{IJK}W^\a_J\OBL_K-2(\g^a)^{\a\b}\O_\b(\N_a\OBL_I+F_{Ia}{}^b\OBL_b) .} } 

In this section we have studied the possibility of having  chain of superfields in curved background similar to $(A_\a,A_m,W^\a)$ of $N=1$ flat superspace in ten dimensions or the corresponding superfields of $N=1$ flat heterotic superspace in ten dimensions. Starting from $\AB_{\a a}$ and $\AB_{\a I}$ which are constrained such that the unintegrated vertex operator \Un\ is in the cohomology of the pure spinor BRST operator. This implies the existence of $\AB_{\b a}$ and $\AB_{aI}$ according to \DAs. The next superfields are the gravitino and gaugino superfields $\WB$ of \Wc\ and \Wci. Note that $\WB^\a{}_a$ satisfies the equation of motion \NWc\ and the  gauge invariance \dWs\ which indicates that this superfield contains a gravitino-like state. In the next section we find the integrated vertex operator of the pure spinor using the results of this section. 

\newsec{The construction of the integrated vertex operator}

As in flat space-time, the integrated vertex operator is defined from the unintegrated vertex operator $U$ by first finding $W$ and $\Wb$ satisfying $QW=\p U, Q\Wb=\pb U$, and then the unintegrated vertex operator, $V$, is given by $QV=\p\Wb-\pb W$. $W$ is given by
\eqn\Whetc{\eqalign{&W= [ \Pi^A\AB_{Aa} + d_\a ( \WB^\a{}_a+2(\g^b)^{\a\b}\O_\b \AB_{ba} ) + J \O_\a \WB^\a{}_a \cr +&\ha N^{cb} ( \FB_{cba} +T_{cb}{}^A \AB_{Aa} - T_{a[c}{}^d \AB_{b]d} + 4 \tau_{cb}{}^d \AB_{da} + 2(\O\g_{cb}\WB_a) )  ] c\Pib^a \cr &  +[ \Pi^A \AB_{AI} + d_\a ( \WB^\a{}_I+2(\g^a)^{\a\b} \O_\b \AB_{aI} ) + J\O_\a\WB^\a{}_I \cr & +\ha N^{ab} ( \FB_{abI} + T_{ab}{}^A \AB_{AI} - F_{I[a}{}^c \AB_{b]c} + 4 \tau_{ab}{}^c\AB_{cI} + 2(\O\g_{ab}\WB_I) ) ] c \Jb^I,  }}
where $N^{ab}=\ha(\l\g^{ab}\o)$ and $J=\l^\a\o_\b$. The calculation of $QW=\p U$ involves the BRST transformation in \brst\ and the use of the equations satisfied by the $\AB$'s and the $\WB$'s derived in the previous section and the Bianchi identities for the torsion, the curvature and the field-strength. The calculation goes like this. When $Q$ acts on $\Pi^\a$  in the term $\Pi^\a (\AB_{\a a} c\Pib^a + \AB_{\a I} c\Jb^I)$ of $W$ will give $\N\l^\a (\AB_{\a a} c\Pib^a + \AB_{\a I} c\Jb^I)$ which will produce $\N U$ after using the Leibniz rule. In the remaining expression, the equations of motion for $\Pib^a$ and $\Jb^I$ are needed. They can be read from \ChandiaHN. The left over term will mix with the remaining terms from $QW$. For example, the expression involving $\Pi^\a \Pib^a$ contains the factor equal to the first equation in \DAs. Similarly for the expression involving $\Pi^\a \Jb^I$ and the second equation in \DAs. We can continue the calculation to finally obtain that $QW=\N U=\p U$.  For $\Wb$ is much easier to obtain and it is given by
\eqn\Wbhetc{\Wb=-\l^\a(\AB_{\a a}\Pib^a + \AB_{\a I} \Jb^I) .}
The pure spinor part of the BRST charge gives zero when it acts on $\Wb$ because \DAs. The remaining terms in the BRST charge contains the right-moving energy-stress tensor which helps to produce $Q\Wb=\pb U$. 

As it was indicated above the integrated vertex operator satisfies $QV=\p\Wb-\pb W$. As in the discussion at the end of section 2, the presence of the right-moving parametrization ghosts allows to find the integrated vertex operator. Because, $W$ has the form $W=cX$ and noting that the unintegrated vertex operator $U$ is given by $U=c\Wb$ we can deduce that $X=V$, the integrated vertex operator. That is
\eqn\intV{\eqalign{&V= [ \Pi^A\AB_{Aa} + d_\a ( \WB^\a{}_a+2(\g^b)^{\a\b}\O_\b \AB_{ba} ) + J \O_\a \WB^\a{}_a \cr +&\ha N^{cb} ( \FB_{cba} +T_{cb}{}^A \AB_{Aa} - T_{a[c}{}^d \AB_{b]d} + 4 \tau_{cb}{}^d \AB_{da} + 2(\O\g_{cb}\WB_a) )  ] \Pib^a \cr & + [ \Pi^A \AB_{AI} + d_\a ( \WB^\a{}_I+2(\g^a)^{\a\b} \O_\b \AB_{aI} ) + J\O_\a\WB^\a{}_I \cr & +\ha N^{ab} ( \FB_{abI} + T_{ab}{}^A \AB_{AI} - F_{I[a}{}^c \AB_{b]c} + 4 \tau_{ab}{}^c\AB_{cI} + 2(\O\g_{ab}\WB_I) ) ] \Jb^I .}}

The action \Shet\ plus $\int d^2z~V$ can be written in the form \Shet\ for redefined background superfields. This is easier to see if one uses superspace coordinates $Z^M$. In this case, the action takes the form 
\eqn\sugra{\eqalign{S'=\int d^2z~&\ha\p Z^M\pb Z^N(G'_{NM}+B'_{NM})+J\pb Z^M\O'_M+\ha N^{ab}\pb Z^M\O'_{Mab} \cr &+d_\a\pb Z^M E'_M{}^\a + \Jb^I \p Z^M A'_{MI}+d_\a\Jb^I {W'}_I^\a+J\Jb^I U'_I+\ha N^{ab}\Jb^I U'_{Iab}\cr &+\o_\a\pb\l^\a+\r_{\cal A}\p\r_{\cal A}  ,\cr}}
which has the form of \Shet\ by removing the primes. Recall that the supermetric $G_{NM}=E_N{}^a E_M{}^b\eta_{ab}$ where $E$ is the vielbein and the other superfields are changed from $M$ type of indices to $A$ type of indices through the vielbein. The relation between primed and unprimed background superfields are
\eqn\prupr{\eqalign{&G'_{NM}=G_{NM}+E_{(N}{}^a\AB_{M)a} ,\quad B'_{NM}=B_{NM}+E_{[N}{}^a\AB_{M]a},\cr &\O_M'=\O_M+E_M{}^a\O_\a\WB^\a{}_a,\cr &\O'_{Mab}=\O_{Mab}+E_M{}^c\left(\FB_{abc}+T_{ab}{}^A\AB_{Ac}-T_{c[a}{}^d\AB_{b]d}+4\tau_{ab}^d\AB_{dc}+2(\O\g_{ab}\WB_c)\right),\cr &E'_M{}^\a=E_M{}^\a+E_M{}^a\left(\WB^\a{}_a+2(\g^b)^{\a\b}\O_\b\AB_{ba}\right),\quad A'_{MI}=A_{MI}+\AB_{MI} ,\cr &W'^\a_I=W^\a_I+\left(\WB^\a{}_I+2(\g^a)^{\a\b}\O_\b\AB_{aI}\right),\cr &U'_I=U_I+\O_\a\WB^\a{}_I ,\cr &U'_{Iab}=U_{Iab}+\left(\FB_{abI}+T_{ab}{}^A\AB_{AI}-F_{I[a}{}^c\AB_{b]c}+4\tau_{ab}{}^c\AB_{cI}+2(\O\g_{ab}\WB_I)\right) .\cr}}
This proves that a supergravity background plus its BRST fluctuations is also a supergravity background. This concludes the construction of the integrated vertex operator for the heterotic string in the pure spinor formulation and in a generic curved supergravity background.

\newsec{Concluding remarks}

In this paper we have generalized the construction of the integrated vertex operator for the heterotic pure spinor string in a flat background to the on-shell supergravity background. The presence of the parametrization ghosts for the right-moving part of the superstring modes are crucial to get the final form the the integrated vertex operator \intV. It would be interesting to study the analogous to the massive vertex operator in flat space of  in curved background. This construction could be difficult because the flat space case is not straightforward \BerkovitsQX.  An application of the vertices studied in this paper is the computation of tree-level scattering amplitudes. That is, the curved background version of the flat space amplitudes of \BerkovitsPH. 

One possible generalization is to find the corresponding results for the type II superstring. The calculation will be more involving because the absence of the $(b,c)$ ghosts. However, the special cases of the superstring in a $AdS_5\times S^5$ background \ChandiaKJA, \ChandiaAFC\ and its plane wave limit \ChandiaEKC\ were already studied, so the general background case can be performed.

\bigskip
\bigskip 
\noindent
{\bf Acknowledgements:} I would like to thank Brenno Carlini Vallilo for useful comments and suggestions. This work is partially supported by FONDECYT grant 1151409.

\listrefs
 
\end